\documentclass[twocolumn,preprintnumbers,a4,pra]{revtex4}
\usepackage{amssymb,amsbsy,amsfonts,amsmath}
\usepackage{epsfig,color}
\usepackage[citecolor=blue,colorlinks=true,linkcolor=blue,urlcolor=blue]{hyperref}
\begin{document}
\title{Deuterium-tritium fusion process in strong laser fields: Semiclassical simulation }
\author{Shiwei Liu $^{1}$, Hao Duan$^{2}$, Difa Ye$^{2}$}
\author{Jie Liu$^{3,4}$}
\email[Email:]{jliu@gscaep.ac.cn}
\affiliation{$^{1}$ Beijing Computational Science Research Center, Beijing 100193, People's Republic of China
\\$^{2}$ Laboratory of Computational Physics, Institute of Applied Physics and Computational Mathematics, Beijing 100088, People's Republic of China
\\$^{3}$ Graduate School, China Academy of Engineering Physics, Beijing 100193, People's Republic of China
\\$^{4}$ CAPT, HEDPS, and IFSA Collaborative Innovation Center of MoE, Peking University, Beijing 100871, People's Republic of China}
\begin{abstract}
In this paper, we investigate the deuterium-tritium (DT) fusion process in the presence of strong laser fields with a semiclassical (SC) method. In this model, two nuclei with a given incident kinetic energy that closely approach each other are simulated by tracing the classical Newtonian trajectories in the combined Coulomb repulsive potentials and laser fields. At the nearest position or classical turning point, quantum tunneling through the Coulomb barrier emerges, and its penetrability is estimated with the Wentzel-Kramers-Brillouin formula. Nuclear fusion occurs after the tunneling, and the total fusion cross section takes the Gamow form. We find that the tunneling penetrability can be enhanced dramatically because the nuclei can closely approach each other due to the quiver motion of the charged nuclei driven by the intense laser fields. We then calculate the DT fusion section for a wide range of laser parameters according to various incident nuclei kinetic energies and obtain the phase diagrams for the enhanced DT fusion. We compare our SC results with the quantum results of the Kramers-Henneberger approximation and the Volkov state approximation.

\end{abstract}
\maketitle
\section{Introduction}

With the advance of chirped pulse amplification techniques \cite{CPA}, modern laser systems can generate intense laser pulses with extension of the intensity range from 10$^{13}$ W/cm$^{2}$ to 10$^{22}$ W/cm$^{2}$. These lasers not only can be applied to the ionization of atoms and molecules \cite{Joachainbook,BeckerRMP2012,liubook} and the acceleration of charged particles \cite{Nat1,Nat2,Nat3} but also provide a potential way to manipulate the nuclear processes. For example, the strong laser fields might increase the $\alpha$ decay rates of heavy nuclear processes by modifying the potential barrier, as reported in recent theoretical works \cite{Casta,Misicu,Delion,Ghinescu,Palffy,Qi,Qi1}. Moreover, important effects of the laser field on deuterium-tritium (DT) fusion processes have attracted wide attention. In the high-laser-frequencies regime, by using a Floquet scattering theory method, \cite{Floquet} showed that the Coulomb barrier penetrability can be increased. Based on the Kramers-Henneberger (KH) approximation, Ref. \cite{Lv} predicts that the fusion cross section is enhanced by several times in magnitude due to the Coulomb barrier suppression effect. The quantum Volkov state approximation (VSA) theory in Ref. \cite{Wang} suggests that the enhancement in the DT section can also emerge in the near-infrared regime because of the increase in the effective nuclei kinetic energy.

In this work, we investigate the DT fusion processes with a semiclassical (SC) method. The fusion reaction can be divided into three processes \cite{Stefanobook}. First, two nuclei approach each other up to a classical turning point. Then, each nucleus tunnels through the Coulomb barrier. Finally, fusion occurs when the nuclei come into contact. In the first step, because the de Broglie wavelength of a particle is much smaller than the interaction length, the relative motion between two nuclei in the combined laser and Coulomb fields can be regarded as classical; thus, the canonical equations can be exploited to determine the nearest position. At the turning point of the Coulomb barrier, quantum tunneling emerges, the penetrability of which can be calculated according to the Wentzel-Kramers-Brillouin (WKB) approximation \cite{Landaubook}. Then, total fusion cross sections are obtained by using the Gamow formula \cite{Gamow,Burbidge}. The SC approach, which describes well the ultra-intense laser-assisted nuclei collision process, is applicable to both high-frequency and low-frequency fields. With the SC approach, we calculate the DT fusion section for a wide range of laser field amplitudes and frequencies with respect to various incident nuclei kinetic energies. We find that the DT fusion probability can be enhanced dramatically in a certain region of the laser parameters that depends on the incident kinetic energy.

The paper is organized as follows. Sec. II presents the model. In Sec. III, we discuss the mechanism of the intense laser-assisted DT fusion process. Sec. IV gives the fusion cross sections in the presence of intense laser fields. Finally, in Sec. V, we present our conclusions.

\section{Model}

The two-body Hamiltonian of DT fusion in the presence of laser fields under the velocity gauge is given by
\begin{eqnarray}
H\left(t\right)&=&\frac{\left[\textbf{p}_{1}-q_{1}\textbf{A}\left(t\right)\right]^{2}}{2m_{1}}
+\frac{\left[\textbf{p}_{2}-q_{2}\textbf{A}\left(t\right)\right]^{2}}{2m_{2}}\nonumber\\
&&+V\left(\textbf{r}_{1}-\textbf{r}_{2}\right),
\label{H2}
\end{eqnarray}
where $m_{1(2)}$ and $q_{1(2)}=e$ are the masses and charges of the deuterium(tritium) nuclei, respectively, specified by $\textbf{r}_{1(2)}$ and $\textbf{p}_{1(2)}$, which are the position vectors and canonical momentum, respectively. $\textbf{A}(t)$ represents the vector potential of the laser field, which is independent of $\textbf{r}$ under the dipole approximation. The two-body Hamiltonian can be divided into two parts $H_{\mathrm{ce}}(t)=[\textbf{P}-Q\textbf{A}(t)]^{2}/2M$ and $H_{\mathrm{re}}(t)=[\textbf{p}-q\textbf{A}(t)]^{2}/2m+V(\textbf{r})$, in which $H_{\mathrm{ce}}$ represents the centroid motion Hamiltonian and $H_{\mathrm{re}}$ represents the relative motion Hamiltonian. $Q=q_{1}+q_{2}$ and $q=(q_{1}m_{2}-q_{2}m_{1})/(m_{1}+m_{2})$ are the centroid motion charge and relative motion charge, respectively. $M=m_{1}+m_{2}$ and $m=m_{1}m_{2}/(m_{1}+m_{2})$ are the total and reduced masses, respectively. $\textbf{P}=M \dot{\textbf{R}}$, $\textbf{p}=m \dot{\textbf{r}}$, $\textbf{R}=(m_{1} \textbf{r}_{1}+m_{2} \textbf{r}_{2})/(m_{1}+m_{2})$ and $\textbf{r}=\textbf{r}_{1}-\textbf{r}_{2}$. $V(\textbf{r})$ is the interaction potential between the DT nuclei, including a short range attractive nuclear potential and a long range Coulomb repulsive potential. The centroid and relative motions can be separated so that the fusion process is determined by the relative motion Hamiltonian:
\begin{eqnarray}
H_{\mathrm{re}}\left(t\right)&=&\frac{\left[\textbf{p}-q\textbf{A}\left(t\right)\right]^{2}}{2m}+V\left(\textbf{r}\right)\nonumber\\
&=&\frac{\left[\textbf{p}-q\textbf{A}\left(t\right)\right]^{2}}{2m}-\Theta\left(\textbf{r}_{\mathrm{n}}-\textbf{r}\right)U_{0}\nonumber\\
&&+\Theta\left(\textbf{r}-\textbf{r}_{\mathrm{n}}\right)\frac{e^{2}}{4\pi\epsilon_{0}|\textbf{r}|},
\label{Hre}
\end{eqnarray}
where the Heaviside function $\Theta\left(r\right)=1$ for $r>0$ and $\Theta\left(r\right)=0$ for $r\leq0$. $\epsilon_{0}$ is the vacuum dielectric constant. For a DT collision, the attractive nuclear potential well $U_{0}$ is approximately $30-40$ MeV, the barrier peak energy is $V_{\mathrm{b}}=370$ keV, and the contact radius $r_{\mathrm{n}}$ is approximately $3.89$ fm.

\begin{figure}[t]
\centering
\includegraphics[width=0.9\columnwidth]{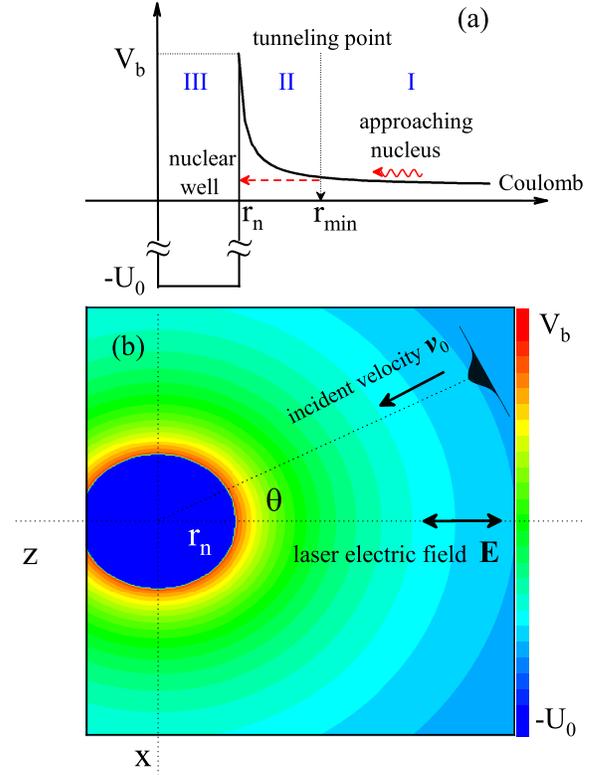}
\caption{(Color online) (a) The process of DT fusion is divided into three regions. Region I: classical motion; Region II: quantum tunneling; Region III: nuclear fusion. (b) Sketch plot of the classical Gaussian wave packet simulations of the DT fusion process.}
\label{fig1}
\end{figure}

As sketched in Fig. \ref{fig1}(a), the relative motion $H_{\mathrm{re}}$ can be divided into three different regions.

In region I, the de Broglie wavelength of the particle (i.e., $\lambda=2\pi\hbar/\sqrt{2 m \epsilon}$, where $\hbar$ is the reduced Planck constant) is much smaller than the interaction length (which can be estimated by suppose the interaction is larger than $1\%$ Coulomb potential) when the incident relative kinetic energy $\epsilon$ is much less than $30$ MeV. This condition is fully satisfied in our following calculations. Therefore, the relative motions of a nucleus can be approximately depicted by the following classical canonical equations:
\begin{eqnarray}
\frac{d\textbf{r}}{dt}=\frac{\partial H_{\mathrm{re}}}{\partial \textbf{p}},\ \ \frac{d\textbf{p}}{dt}=-\frac{\partial H_{\mathrm{re}}}{\partial \textbf{r}}.
\label{ce}
\end{eqnarray}
The vector potential of a linear polarization laser field can be specified as $\textbf{A}(t)=A_{0}f(t)\cos(\omega t+\phi_{0})\hat{e}_{z}$, where $\omega$ is the laser frequency and $\phi_{0}$ is the initial phase. The magnitude of the amplitude $A_{0}=\sqrt{2I/\epsilon_{0}c}/\omega$, where $c$ is the speed of light in a vacuum and $I$ is the laser intensity. $f(t)$ denotes the field envelope:
\begin{equation}\label{field}
f(t)=\left\{\begin{array}{ll}
\sin ^{2}\left(\frac{\omega t}{12}\right) & \text { if } 0 \leqslant t \leqslant 6\pi /\omega, \\
1 & \text { if } t > 6\pi /\omega.
\end{array}\right.
\end{equation}
\begin{figure*}[t]
\centering
\includegraphics[width=2\columnwidth]{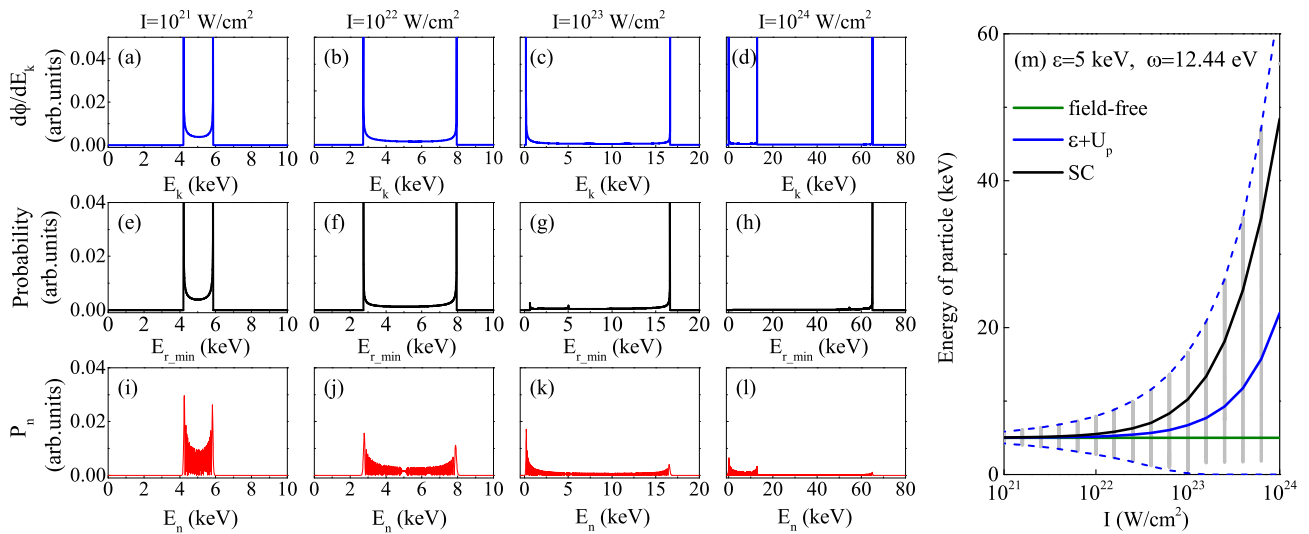}
\caption{(Color online) (a)-(l) The energy distribution of particles for different laser intensities: the analysis of Eq. \eqref{traE} (blue line), the SC results (black line) and the VSA results (red line). The frequency of the laser field is $\omega=12.44$ eV. (m) Distribution on energies $E_{r_{-}\min}=e^{2}/(4\pi\epsilon_{0}r_{\mathrm{min}})$ (gray dotted lines) calculated from the SC model with respect to various laser intensities. The average $E_{r_{-}\min}$ over one laser cycle (black line) and the value of $\epsilon+U_{p}$ (blue line) are also plotted for comparison. The blue dashed lines represent the boundaries predicted by Eq. \eqref{traE0}. The initial incident kinetic energy is $\epsilon=5$ keV (olive line).  }
\label{fig2}
\end{figure*}
The initial state of the incident nucleus is described by a Gaussian wave packet, as sketched in Fig. \ref{fig1}(b), satisfying the following distribution function:
\begin{equation}
F(\textbf{r}, \sigma_{\mathrm{r}})=\frac{1}{\sqrt{2 \pi} \sigma_{\mathrm{r}}} \exp \left[-\frac{(\textbf{r}-\textbf{r}_{0})^{2}}{2 \sigma_{\mathrm{r}}^{2}}\right],
\label{Gaussian}
\end{equation}
where $\textbf{r}_{0}$ is the center position of the incident nucleus and the width of the wave packet depends on the initial kinetic energy $\epsilon$ through the relation $\sigma_{\mathrm{r}}=\lambda$ \cite{Heller,Martinez,Belkacem,Meckel,Guzman,Liu}.

In region II, i.e., between the classical turning point $r_{\mathrm{min}}$ and contact point $r_{\mathrm{n}}$, quantum tunneling occurs, and the corresponding penetrability can be described by the WKB approximation \cite{Landaubook}:
\begin{eqnarray}
P\left(\epsilon\right)=\mathrm{exp}\left\{{-\frac{2}{\hbar}\int_{r_{\mathrm{n}}}^{r_{\mathrm{min}}}\sqrt{2m\left[V(r)-V\left(r_{\mathrm{min}}\right)\right]}dr}\right\}.
\label{P}
\end{eqnarray}
In the above formula, the turning point $r_{\mathrm{min}}$ can be numerically obtained by solving canonical equations (\ref{ce}) with the standard 4-5th Runge-Kutta algorithm.

In region III, the nuclear fusion process is described by the so-called astrophysics factor $S$ \cite{Gamow,Stefanobook}:
\begin{eqnarray}
S\left(\epsilon\right)&=&a+\frac{b}{\pi}\frac{d}{4\left(\epsilon-f\right)^{2}+d^{2}},
\label{Sfact}
\end{eqnarray}
with the fitting parameters $a$, $b$, $d$ and $f$. By employing the Gamow form, we obtain the fusion cross section by
\begin{eqnarray}
\sigma\left(\epsilon\right)&=&\frac{S\left(\epsilon\right)}{\epsilon}P\left(\epsilon\right),
\label{sigma}
\end{eqnarray}
which is a simple product of the penetrability and $S$ factor and geometry cross section $1/\epsilon$.

\section{Fusion perspective on a DT collision in a strong laser field}

In this section, we first restrict our discussion to a one-dimensional situation to achieve insight into the mechanisms of the laser-assisted DT fusion process.

\subsection{Particle energy distribution }

Neglecting the influence of the Coulomb potential, the particle motion is given by
\begin{equation}\label{trav}
\left\{\begin{array}{l}
v(t)=v_{0}-v_{\mathrm{quiver}} \cos (\omega t + \phi_{0}), \\
r(t)=r_{0}+v_{0} t-r_{\mathrm{quiver}} \sin (\omega t + \phi_{0}),
\end{array}\right.
\end{equation}
where $v_{\mathrm{quiver}}= q E_{0} / m \omega $ and $r_{\mathrm{quiver}}= q E_{0} / m \omega^{2} $ are the quiver velocity and the quiver distance in the laser field, respectively. $E_{0}=\sqrt{2I/\epsilon_{0}c}$ is the magnitude of the laser electric field. The particle energy in the laser field takes the form
\begin{equation}\label{traE0}
E_{k} =\frac{1}{2}m\left(v_{0}-v_{\mathrm{quiver}} \cos \phi_0\right)^{2},
\end{equation}
where $\phi_{0}$ is the initial phase. Then, the distribution of the kinetic energy of the particles in the laser fields can be evaluated by the following expression:
\begin{equation}\label{traE}
P\left(E_{k}\right) \mathrm{d} E_{k}=\Gamma(\phi_{0}) \mathrm{d} \phi_{0}, \quad P\left(E_{k}\right)=\Gamma(\phi_{0}) \frac{\mathrm{d} \phi_{0}}{\mathrm{d} E_{k}},
\end{equation}
where $\Gamma(\phi_{0})=1$ represents the uniform distribution of $\phi_{0}$.

The plots of the distribution of particle energy $E_{k}$ for various laser intensities are shown in Figs. \ref{fig2}(a)-(d). For a given laser field parameter that satisfies $v_{\mathrm{quiver}} < v_{0}$, the particle energies are distributed in an interval where the maximum value is $m\left(v_{0}+v_{\mathrm{quiver}}\right)^{2}/2$ and the minimum value is $m\left(v_{0}-v_{\mathrm{quiver}}\right)^{2}/2$ according to Eq. \eqref{traE0}. With increasing laser intensities, the range of the nonzero energy distribution becomes wider due to the increase in quiver velocity $v_{\mathrm{quiver}}$, as shown in Figs. \ref{fig2}(a)-(c). When the laser field is large enough that $v_{\mathrm{quiver}}>v_{0}$ , as shown in Fig. \ref{fig2}(d), the particle energy distribution has three peaks, located at $E_{k}=0$, $E_{k}=m\left(v_{0}-v_{\mathrm{quiver}}\right)^{2}/2$ and $E_{k}=m\left(v_{0}+v_{\mathrm{quiver}}\right)^{2}/2$.
In Fig. \ref{fig2}(m), the blue dashed lines represent the energy boundaries predicted simply by Eq. \eqref{traE0}.

\begin{figure}[pt]
\centering
\includegraphics[width=\columnwidth]{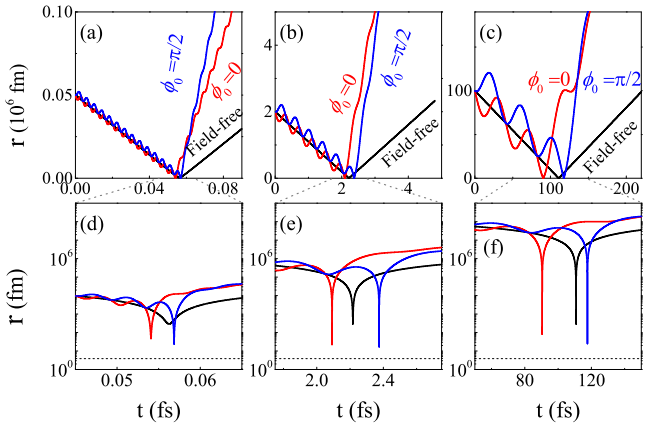}
\caption{(Color online) Typical classical trajectories of a DT collision driven by a laser field for (a) $\omega=1000$ eV, (b) $\omega=10$ eV and (c) $\omega=0.1$ eV, with different initial laser phases $\phi_{0}$. The laser intensities are $10^{28}$ W/cm$^{2}$, $10^{24}$ W/cm$^{2}$ and $10^{20}$ W/cm$^{2}$. The initial incident kinetic energy is $\epsilon=5$ keV. (d)-(f) are enlarged views of (a)-(c). The gray dashed lines represent the contact radius $r_{\mathrm{n}}$.}
\label{fig3}
\end{figure}

Taking the Coulomb interaction between nuclei into account, the motions of a particle cannot be solved analytically. We exploit the standard 4-5th Runge-Kutta algorithm to solve the coupled classical equations (\ref{ce}) numerically; the typical trajectories of a laser-assisted DT collision are shown in Fig. \ref{fig3}. When the DT nuclei are far away from each other so that the Coulomb potential between them is much less than the laser field, the laser dominates the particle motion, and the corresponding classical trajectory is a combination of the quiver and drift motions. The oscillating motions in Figs. \ref{fig3}(a)-(c) can be approximately described by Eq. \eqref{trav}. When the nuclei are closely approaching each other, the particle performs a reciprocating motion due to the interplay between the Coulomb repulsive potential and laser field driving and is then bounced back at the classical turning point $r_{\mathrm{min}}$, as shown in Fig. \ref{fig3}(d)-(f). The position of the turning point $r_{\mathrm{min}}$ depends on the laser frequency and intensity $I$ as well as the initial phase. Compared to that of the field-free case, we find that $r_{\mathrm{min}}$ can be much smaller, indicating that two nuclei approach each other much more with the assistance of intense laser fields. We denote the particle energy at this position as $E_{r_{-}\min}=e^{2}/(4\pi\epsilon_{0}r_{\mathrm{min}})$, and in Figs. \ref{fig2}(e)-(h), we plot its distributions with respect to various laser intensities.

When the intensity of the laser field is relatively smaller ($I$=$10^{21}$ and $10^{22}$ W/cm$^{2}$ in Fig. \ref{fig2}), the range or boundaries of the nonzero energy distribution from our SC results are almost the same as those of the classical analysis of Eq. \eqref{traE0}. With increasing intensity, the energy distributions change substantially. However, the upper boundaries of the energy distribution of the SC results can still be predicted approximately by Eq. \eqref{traE0}, while the lower boundaries cannot. To demonstrate these changes more clearly, we take the average of the energy $E_{k}$ and $E_{r_{-}\min}$ over one laser cycle, as shown in Fig. \ref{fig2}(m), for comparison. The figure demonstrates that laser-field-assisted collision energy with the Coulomb potential can be much larger than $\epsilon+U_{p}$. Here, $U_{p}=(q E_{0})^{2} / (4m \omega^{2})$ is termed ponderomotive energy, donating the average quiver energy of a charged particle in a laser field.

For comparison, we also show the calculations of the VSA results \cite{Wang} in Figs. \ref{fig2}(i)-(l). Without considering the Coulomb effects, VSA theory describes the motion of a charged particle in a laser field as a Volkov state \cite{Volkov}:
\begin{equation}\label{Volkov0}
\begin{aligned}
\psi_{V}\left(\textbf{r}, t\right)&=\frac{1}{\left(2 \pi\hbar\right)^{3 / 2}} e^{\frac{i}{\hbar}\textbf{p} \cdot \textbf{r}- \frac{i}{2m\hbar}\int_{0}^{t} \left[\textbf{p}-q\textbf{A}\left(\tau\right)\right]^{2}\mathrm{d} \tau} .\\.
\end{aligned}
\end{equation}
Therefore, the particle energy can be determined by absorbing or emitting integer numbers of photons in the laser field, i.e., $E_{\mathrm{n}}=\epsilon+U_{\mathrm{p}}+n\omega$. The probability of energy $E_{\mathrm{n}}$ is obtained by carrying out the Fourier expansion of Eq. \eqref{Volkov0}:
\begin{equation}\label{Volkov1}
\begin{aligned}
\mathcal{F}[\psi_{V}(\textbf{r}, t)]&=\frac{1}{\left(2 \pi\hbar\right)^{3 / 2}} \exp \left(\frac{i}{\hbar} \textbf{p} \cdot \textbf{r}\right) \\
&\sum_{n=-\infty}^{\infty}e^{-\frac{i u}{\hbar}} \left[(-1)^{n}iJ_{n}(u, v)\right]  e^{-\frac{i}{\hbar}\left(\epsilon+U_{p}\right)t-in \omega t},
\end{aligned}
\end{equation}
$P_{\mathrm{n}}=\left|(-1)^{n}iJ_{n}(u, v)\right|^{2}$, where $J_{n}(u, v)$ is the generalized Bessel function \cite{Reiss}, $u=(q p A_{0} ) / (m\hbar \omega)$ and $v=(q^{2} A_{0}^{2}) / (8 m\hbar \omega)$. Compared with Figs. \ref{fig2}(a)-(b) and (e)-(f), the VSA (Figs. \ref{fig2}(i)-(j)) shows similar energy distributions, except for some rapid oscillations due to quantum interference. In addition, the VSA results are quite different from the SC results, especially for relatively strong laser fields, as clearly seen by comparing Figs. \ref{fig2}(h) and (l), which is mainly because the VSA completely ignores the influence of the Coulomb potential.

\subsection{Penetrability at $r_{\mathrm{min}}$}

\begin{figure}[pt]
\centering
\includegraphics[width=\columnwidth]{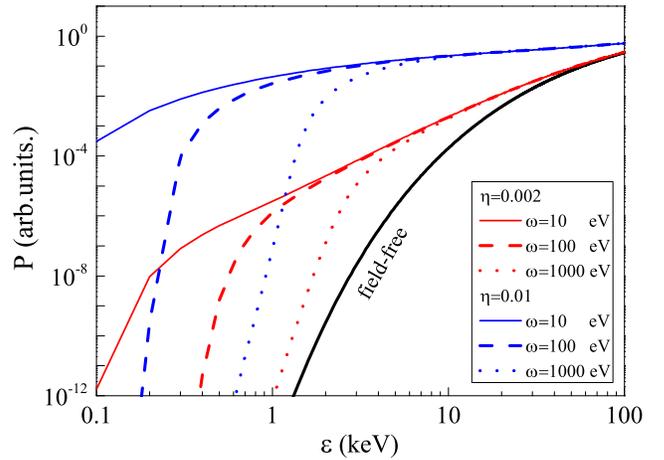}
\caption{(Color online) The penetrability P vs initial energy $\epsilon$ for $\eta=0.01$ (blue) and $\eta=0.002$ (red) at different frequencies. The black lines represent the situation without a laser field.}
\label{fig4}
\end{figure}

The minimum position or the turning point of classical trajectories $r_{\mathrm{min}}$ depends on the laser parameters, incident energy and initial phase. Therefore, the penetrability expressed by Eq. \eqref{P} is also a function of multiple variables. We average $P(\epsilon)$ over the initial phases and demonstrate in Fig. \ref{fig4} the average probability versus the incident kinetic energy, laser field frequency and scaled quiver velocity $\eta=v_{\mathrm{quiver}}/c$.

We find that for a fixed laser field frequency, with increasing $\eta$, the penetrability can be increased considerably. Taking $\epsilon=1$ keV and $\omega=10$ eV as an example, for $\eta=0.002$ and 0.01, the enhancements are over eight orders and twelve orders of magnitude, respectively. For the same quiver velocity $\eta$, a lower-frequency laser field will lead to more effective enhancement in penetrability than will high-frequency fields, especially for the regime of relatively smaller incident energies.

The enhancement of the penetrability is due to the decrease in $r_{\mathrm{min}}$.
In the absence of a laser field, the incident particle reaches the classical turning point when its initial kinetic energy is completely converted into the Coulomb repulsive potential. In the presence of a laser field, the nuclei subjoin a quiver motion and lead to a nearer distance between nuclei that can increase the penetrability dramatically, since the  tunneling probability usually depends sensitively on the barrier width or height with an exponential form. For a relatively smaller incident energy, i.e., $\epsilon<$10 keV, the turning points in the absence of laser fields is a little farther away from the origin, and the effects due to the quiver motion are expected to be more significant. For a smaller field frequency, the distance of the quiver motion increases, which can also lead to a more significant enhancement in the penetrability.

\section{The cross section of DT fusion in strong laser fields}

In this section, we attempt to calculate the DT fusion cross section quantitatively with our three-dimensional SC approach. In practical calculations, the incident nucleus is approximately described by a Gaussian wave packet whose width depends on $\epsilon$ according to Eq. \eqref{Gaussian}. We can obtain the initial position distribution of the incident nucleus via the Monte Carlo sampling technique \cite{liubook}. More than $10^{5}$ classical trajectories are used for statistics to guarantee numerical convergence.

\begin{figure}[pt]
\centering
\includegraphics[width=\columnwidth]{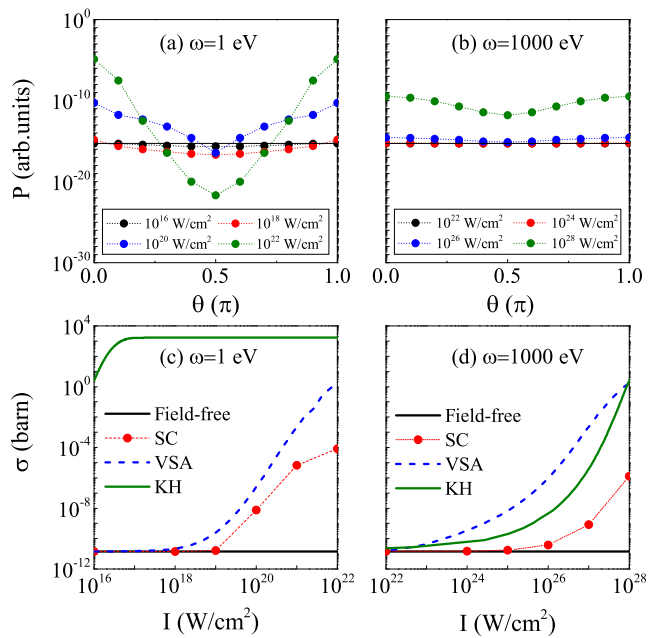}
\caption{(Color online) (a)-(b) Angular dependence of the penetrability for various laser intensities and frequencies. The horizontal line in each figure indicates the field-free values. (c)-(d) Total fusion cross section with various laser intensities and frequencies, obtained by the SC method (red solid dots). Other theories are presented for comparison: the KH results \cite{Lv} (olive lines) and the VSA results \cite{Wang} (blue dashed lines). The initial incident energy $\epsilon=1$ keV.}
\label{fig5}
\end{figure}

In Figs. \ref{fig5}(a)-(b), we show the angular dependence of the penetrability for various laser intensities and frequencies. Here, $\theta$ is the angle between the incident velocity and the electric field direction; see Fig. \ref{fig1}(b) for the configuration. The penetrability reaches its maximum when the incident direction of the nuclei is parallel to the laser field polarization direction ($\theta=0$ or $\pi$). When the incident direction is perpendicular to the field direction ($\theta=\pi/2$), the penetrability reaches its minimum, the value of which is even smaller than that of the field-free case. This result occurs because the quiver motion driven by a laser field is in the direction perpendicular to the incident velocity direction and therefore cannot effectively reduce the minimum position. Actually, it usually increases the distance between nuclei at the turning points. This effect becomes more apparent with increasing field strength or decreasing field frequency.

The DT fusion cross section $\sigma$ is obtained by the Gamow form of Eq. \eqref{sigma}. The parameters in Eq. \eqref{Sfact} are $a=7153 \mathrm{keV\cdot barn}$, $b=3.899\times10^{7} \mathrm{keV^{2}\cdot barn}$, $d=61.01 \mathrm{keV}$ and $f=52.19 \mathrm{keV}$, which are obtained by fitting the experimental section data with our calculation in the field-free situation \cite{Stefanobook}. The total fusion section is obtained by the angle-averaged scheme:
\begin{equation}\label{sigmaeff}
\sigma(\epsilon)=\frac{1}{2} \int_{0}^{\pi} \frac{S(\epsilon)}{\epsilon}P(\epsilon, \theta) \sin \theta d \theta.
\end{equation}
We plot the total fusion cross section $\sigma(\epsilon)$ of our SC calculation results for various laser intensities and frequencies in Figs. \ref{fig5}(c) and (d). Quantum results obtained with the KH \cite{Lv} and VSA \cite{Wang} methods are also demonstrated for comparison.

Compared with that of the field-free case, we find that the fusion section rapidly increases after $I=10^{19}$ W/cm$^{2}$ for the low-frequency case of $\omega=1$ eV and after $I=10^{25}$ W/cm$^{2}$ for the high-frequency case of $\omega=1$ keV. The enhancement is very laser frequency dependent; for $\omega=1$ eV, it is above approximately six orders of magnitude at $I=10^{22}$ W/cm$^{2}$, which is the maximum intensity in the current intense laser facility of the Ti:sapphire laser \cite{Bahk}.
For $\omega=1$ keV, the enhancement is less than one order of magnitude at $I=10^{26}$ W/cm$^{2}$. For a low frequency, our results qualitatively agree with VSA theory. However, quantitatively, VSA theory overestimates by approximately one order of magnitude at $I=10^{20}$ W/cm$^{2}$ and by approximately five orders of magnitude at $I=10^{22}$ W/cm$^{2}$. For the high-frequency case, where KH theory applies, our results are closer to the KH results but deviate largely from the VSA approximation.

\begin{figure}[pt]
\centering
\includegraphics[width=\columnwidth]{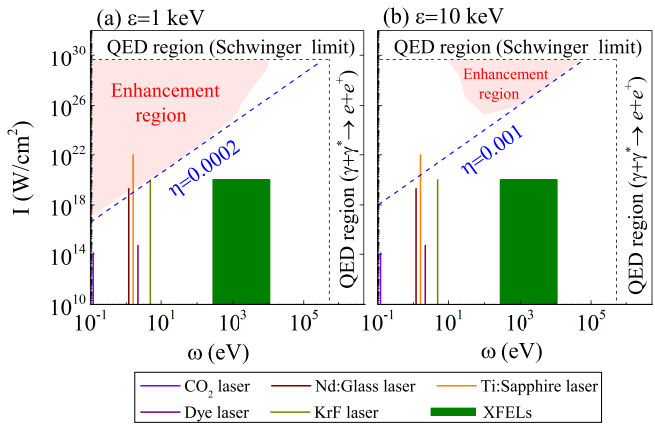}
\caption{(Color online) Phase diagram of the effective region of laser-field-enhanced DT fusion cross sections for different initial energies: (a) $\epsilon=1$ keV and (b) $\epsilon=10$ keV. The typical parameters of various laser facilities at the present time are shown, e.g., CO$_{2}$ laser with a peak intensity of $1.3\times10^{14}$ W/cm$^{2}$ and a frequency of $0.12$ eV \cite{Daido}; Nd:glass laser with a peak intensity of $2\times10^{19}$ W/cm$^{2}$ and a frequency of $1.18$ eV \cite{Badziak}; Ti:sapphire laser with a peak intensity of $10^{22}$ W/cm$^{2}$ and a frequency of $1.55$ eV \cite{Bahk}; dye laser with a peak intensity of $5\times10^{14}$ W/cm$^{2}$ and a frequency of $2.12$ eV \cite{Perry}; KrF laser with a peak intensity of $10^{20}$ W/cm$^{2}$ and a frequency of $5$ eV \cite{Sadighi}; and hard X-ray free-electron laser (XFEL) with a peak intensity of $10^{20}$ W/cm$^{2}$ and variable frequencies from $280$ eV to $11.2$ keV \cite{XFELs,Bucksbaum}. The dashed blue lines denote (a) $\eta=0.0002$ and (b) $\eta=0.001$, respectively. The horizontal black dashed lines and the vertical black dashed lines represent the Schwinger limit of the laser intensity, $4.6\times10^{29}$ W/cm$^{2}$, and the QED limit of the laser frequency, $0.51$ MeV, respectively.}
\label{fig6}
\end{figure}

As shown in Figs. \ref{fig5}(c) and (d), a critical value of the laser intensity exists, beyond which the influence of the lasers becomes significant. In fact, the critical values depend on the field frequencies and the incident energy. We numerically explore a wide range of parameters and show the phase diagram of the effective enhancement region in the parameter plane of the laser frequency and intensity in Figs. \ref{fig6}(a) and (b) for the incident energy $\epsilon=1$ keV and $\epsilon=10$ keV, respectively. In our calculations, the laser intensity is not allowed to exceed the Schwinger limit, i.e., $I_{0}=(m_{e}^{4} c^{5}) / (\mu_{0} e^{2} \hbar^{2})$ \cite{Schwinger}, and the field frequency must be less than $\omega_{0}= m_{e} c^{2} / \hbar$ to avoid the $\gamma+\gamma^{*} \rightarrow e+e^{+}$ process. Here, $\mu_{0}$ is the permeability of a vacuum.

For the incident energy $\epsilon=1$ keV, the region of effective enhancement is mainly in the low-frequency and high-intensity regime, bounded approximately by the scaled quiver velocity $\eta>0.0002$. In contrast, for $\epsilon=10$ keV, the region of effective enhancement substantially shrinks, bounded approximately by the scaled quiver velocity $\eta>0.001$ and $\omega >10$ eV. These phenomena occur because for $\epsilon=10$ keV, the de Broglie wavelength of the initial wave packet is small; thus, nuclei with large incident angles will have less opportunity to collide with each other. This influence is more serious at low frequencies because the relatively larger quiver motion perpendicular to the direction of the incident velocity will lead to dramatic dispersion of the wave packet.

\section{Conclusions}

In summary, we develop an SC approach to achieve insight into the influence of strong linearly polarized laser fields on the barrier penetrability and fusion cross sections of DT fusion. Our model consists of three processes: the particle scattering in the combined repulsive Coulomb potential and laser field, the quantum tunneling process of the Coulomb barrier, and the nuclear fusion. This model is worth comparing with the well-known SC model of the laser-atom interaction \cite{Corkum,liu1997}. The latter model describes  ionization processes of an atom in an intense laser field: a valence electron is first ionized by quantum tunneling through the attractive Coulomb potential barrier of its parent ion and then moves (or scatters) classically in the combined attractive Coulomb potential and laser field. The model is applied to study the nonperturbation phenomena in strong-field physics, such as above-threshold ionization, high-order harmonic generation and nonsequential ionization (see Ref. \cite{liubook} for an example). Both SC models contain quantum tunneling and classical scattering processes, while their time sequences are reversed.

By applying the developed SC approach, we calculate the DT fusion cross sections over a wide range of laser parameters and plot the phase diagrams of the parameter regions in which the laser field can enhance the DT fusion effectively. These predictions require future experimental testing that might be performed in ultra-intense laser facilities such as the Extreme Light Infrastructure (ELI) \cite{ELI} experimental measurements. Our current results are calculated under the nonrelativistic condition. Discussions on the relativistic effects are of interest for future work.

\section*{Acknowledgments}
We are grateful to Dr. Wenjuan Lv for valuable discussions. This work is supported by funding from the National Natural Science Foundation of China (NSFC) (Grants No. 11775030, No. 11822401, No. 12047510, No. U1930402 and No. U1930403).



\end{document}